# Quantum Anticipation Explorer

by Hans-Rudolf Thomann

February 2011

Author address    Rollenbergstrasse 2, CH-8463 Benken ZH, hr@thomannconsulting.ch

            

# CONTENTS

**Abstract**

Quantum anticipation explorer is a computer program allowing the numerical exploration of quantum anticipation which has been analyzed in arXiv:0810.183v1 and arXiv:1003.1090v1 for H-Atom, equidistant, random and custom spectra. This tool determines the anticipation strength at those times where orthogonal evolution is possible. This paper is the user's guide explaining its capabilities, installation and usage, and documenting the mathematics and algorithms implemented in the software. A zip file containing the setup and documentation can be downloaded from http://www.thomannconsulting.ch/public/aboutus/aboutus-en.htm free of cost.

Quantum Anticipation Explorer

# 0 Introduction

Quantum anticipation has been introduced and analyzed in (Thomann, 2008) and (Thomann, 2010) from which we briefly recall the basic terms and facts.

Under a Hamiltonian $H$, a quantum state $q$ evolves into an orbit $q_t = U_t q$, where $U_t = e^{-\frac{iH}{\hbar}t}$ is defined on the closure of $D(H)$. By spectral theory (Reed, 1980), the spectral measure $\mu_q(\lambda)$ of $q$ and $H$ is uniquely defined by $(q, f(H)q) = \int f(H) d\mu_q(\lambda)$ for any analytic function $f$.

For fixed step size $T > 0$, $q_0 = q$ and $n \in \mathbb{Z}$, an evolution of order $L \geq 0$ is given by the primary sequence $q_n = U_T^n q_0$ and the dual sequence, $r_n = U_T^n r_0$, such that $(r_k, q_l) = \delta_{kl}$ ($k, l \in \mathbb{Z}$, $|k - l| \leq L$). The amplitudes $(r_t, q_0)$ are determined by the spectral measure $d\mu_q$. Positive evolutions are defined as those with positive $\rho$.

Every positive evolution contains an *embedded orthogonal evolution*, driving the component $s_0$ of $q_0$ of size $\zeta$ through $L + 1$ mutually orthogonal states. Evolutions (positive and non-positive) of any order exist for point, singular-continuous and absolutely continuous spectrum, respectively.

Quantum-Mechanical anticipation (and retrospection, its time-reverse) is defined for the embedded orthogonal evolution in terms of the anticipation amplitudes $\alpha_n = (s_n, U_\Delta s_0)$, and the anticipation probabilities $p_n = |\alpha_n|^2$, expressing the result of a measurement of $s_0$ at time $\Delta$, where $|n| \leq L$.. Measurements anticipate future (or recall past) states $s_n$ with probability $p_n$, which is the reason for the naming.

While the referenced publications focused on anticipation for fixed step size $T$, Quantum Anticipation Explorer evaluates anticipation for values of $T$ varying in a range and displays graphical and numerical results. It does so starting from a point spectrum which may be the H-atom's, equidistant in an interval, randomly chosen in an interval or custom-defined by the user.

Find this free software at http://www.thomannconsulting.ch/public/aboutus/aboutus-en.htm and start exploring quantum anticipation. As you will see, anticipation occurs in the H-atom and all other types of spectra at many times and varying strengths.

In the first chapter we provide advice for the download, installation and uninstallation of Quantum Anticipation Explorer. The second chapter contains the user's guide. The third chapter specifies all terms found in the user's guide, the operations performed by the software, the mathematical background and the algorithms. The final chapter provides some expamples.

Quantum Anticipation Explorer

# 1 Getting started

## 1.1 System requirements

Quantum anticipation explorer runs on PCs[1] with Microsoft Windows XP, Vista, 7 and above.

## 1.2 Download Quantum Anticipation Explorer

Browse to http://www.thomannconsulting.ch/public/aboutus/aboutus-en.htm and click the link to Quantum Anticipation Explorer. A file download window opens, inviting you to download a zip folder. Store the file on your desktop[2].

## 1.3 Install Quantum Anticipation Explorer

Before installing, uninstall any previously installed version, as explained in 1.5 below.

Open the zip folder by a double-click. It contains three files:

| | |
|---|---|
| Setup.exe | Install starter program |
| QuantumAnticipationExplorer.msi | Microsoft installer |
| UsersGuide.pdf | This document |

Double-click Setup.exe to start the installation[3]. The following prompts appear:

- License agreement for Microsoft .Net Framework 3.5 (click "accept")
- Welcome to the Quantum Anticipation Explorer Setup Wizard (click "next")
- Select Installation Folder. You can specify whether the program is installed for you only (default) or for everyone, accept or change the proposed selection and check the Disc Cost (less than 2Mbytes). Click "next" to continue.
- Confirm Installation. Click "next" to continue.
- Installation Complete. Click "close".

## 1.4 Start Quantum Anticipation Explorer

In the Start\All Programs menu you will now find a new folder QuantumAnticipationExplorer with two entries:

| | |
|---|---|
| QuantumAnticipationExplorer | Launches the program |
| UsersGuide | Opens the pdf file |

## 1.5 Uninstall Quantum Anticipation Explorer

Open control panel, and open the software/programs tab. Right-click QuantumAnticipatio-nExplorer, select "Uninstall" and confirm your selection when prompted. The program and all folders created during install will be removed.

---

[1] Not yet on quantum computers☺
[2] It is best practice to perform a virus check before proceeding.
[3] You need Administrator rights for this operation. Depending on your security settings and operating system, you may be prompted to confirm your intention to install.

Quantum Anticipation Explorer

## 2  User's Guide

Start\All Programs\QuantumAnticipationExplorer to start up Quantum Anticipation Explorer, and then click the "Go" button. The Quantum Anticipation Window opens and shows your first evaluation: The H-atom's anticipation statistics.

This window (see next page) is split into four parts, in top-down order: Numerical output, text box, graphics box and controls section.

The following sub-sections lead you through these four parts with the aim to enable you to use the program and to read its displays. What it does, what the displays are and what's behind all this will be explained in chapter 3.

### 2.1  Controls

The controls are grouped into five panels. The radio buttons in the first panel on the left hand side control the search mode:

| | |
|---|---|
| Continuous | Evaluate the curves selected in steps of "Step size" from "From" to "To". |
| Random | Evaluate the curves selected at randomly selected point, one in each grid cell defined by "Step size", "From" and "To". |
| Seek positive | Starting from "From" in steps of "Step size" seeks the least time $T$ at which a positive measure exists for the spectrum selected, and draw the reduced spectral measure. |
| Seek equal | Starting from "From" in steps of "Step size" seeks the least time $T$ at which the reduced spectrum is equidistant, and draw the reduced spectral measure. |
| Seek dim. chg. | Starting from "From" in steps of "Step size" seeks the least time $T$ at which the problem dimension changes, and draw the reduced spectral measure. |
| Single | Perform a single evaluation at time "From", and draw the reduced spectral measure. |

The second panel specifies the problem:

| | |
|---|---|
| Spectrum | Underlying spectrum:<br>- H Atom: $-2\pi/n^2$ $(n = 1, \ldots, d)$<br>- Equidistant: $2\pi n/d$ $(n = 1, \ldots, d)$<br>- Equidistant alternating: $2\pi(n/d + (n-1) \bmod 2)$ $(n = 1, \ldots, d)$<br>- Random: $d$ random points in $[0, 4\pi)$<br>- Random alternating: $d$ random points, odd numbered lines in $[0, 2\pi)$, even numbered in $[2\pi, 4\pi)$<br>- Prescribed: Defined by input entered into the input text box (see $$$)<br>- Previous: Same spectrum as in previous evaluation. |
| Measure | Method to determine the spectral measure:<br>- Optimum: Maximize anticipation look-ahead<br>- Equal: All spectrum points have equal weight $1/d$<br>- Random: $d$ random values in $[0, 1)$ normalized to unit sum.<br>- Prescribed: Defined by input entered into the input text box (see 2.3)<br>- Previous: Same spectral measure as in previous evaluation. |
| Order | Anticipation order, L |
| Dimension | Dimension, d: Number of eigenvalues, $d \geq 2L + 1$ |
| Location | Location of the measurement point, in $T$ units. |



# Quantum Anticipation Explorer

   7/18

Quantum Anticipation Explorer

The third panel selects the curves to be drawn in continuous and random mode:

| | |
|---|---|
| Anticipation | Anticipation look-ahead |
| Probability | Total anticipation probability |
| Variance | Variance of the reduced spectrum |

The fourth panel groups the time and action settings:

| | |
|---|---|
| Numerator | Numerator. Double-click the label to reset. Double-click into the box to enter Pi. |
| Denominator | Denominator. Double-click the label to reset. |
| From | Start of time interval. Click the label to reset. Double-click into the box to enter Numerator/Denominator. |
| To | End of time interval. Click the label or double-click into the box to reset to 72, the period length of the H-atom at dimension 3. |
| Step size | Step size of successive evaluations. |
| Step number | Display of the step number in continuous and random mode. |
| Time | Display of the time in continuous and random mode. |
| Go | Start the evaluation in continuous, random, seek equal and single mode. |
| Stop | Stop an ongoing evaluation. |
| > | In single or seek mode (except seek equal), move forward. |
| < | In single or seek mode (except seek equal), move backward. |

The right-most panel facilitates the processing of evaluation results:

| | |
|---|---|
| Show Max | After an evaluation in continuous or random mode, displays the time, evaluation results and spectrum with maximum anticipation look-ahead. |
| Save | Save results. See the save window below. |
| Cut | Cuts results to the clip board, from where they can be pasted into documents.<br>- Data checkbox not checked: Screen shot of the whole window.<br>- Data checkbox checked: Contents of the textbox, in text format. |
| Print | Print a hardcopy:<br>- If nothing checked: Whole window<br>- Data checked: Numerical output and textbox<br>- Graphics checked: Graphics<br>- Both checked: Numerical output, textbox and graphics. |

The save box allows you to select the directory and file name for the data saving. The default file name is "Data" in your documents folder. Your selection is remembered and this file name is used for the data file and the image files, as explained below.

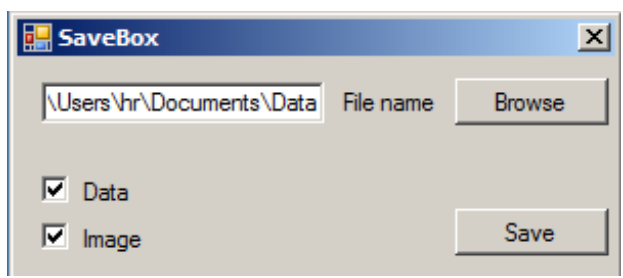

If "Image" is checked, then the graphics are stored as a Jpeg file, with the date and time appended to the file name.

     

Quantum Anticipation Explorer

If "Data" is checked, then the data listed below are appended to a csv-file (comma-separated values file). This file is easily imported into Microsoft Excel and other spread-sheet applications.

| | |
|---|---|
| Date/Time | Date and time of the data saving |
| Search mode | Search mode setting |
| Spectrum type | Spectrum type setting |
| Measure type | Measure type setting |
| Order | Order setting |
| Dimension | Dimension setting |
| Location | Location setting |
| From | "From" value |
| To | "To" setting |
| Number of steps | Number of points evaluated, $[(To - From)/StepSize]$ |
| Non-narrow | Points with non-narrow reduced spectrum[4] |
| Degenerate | Points with degenerate spectrum[4] |
| Singular | Points with singular spectrum or solution[4] |
| Positive | Points with positive solution[5] |
| Zero | Points with non-zero dimension less than minimum dimension[6] |
| Non-zero dim. | Average number of eigenvalues with non-zero ($> 10^{-4}$) measure[7] |
| Max. measure | Peak value of the measure of any eigenvalue |
| Ave. variance | Average variance[8] |
| Ave. probability | Average total anticipation probability[7] |
| Max. probability | Maximum total anticipation probability |
| Ave. Anticipation | Average anticipation look-ahead[7] |
| Max. Anticipation | Maximum anticipation look-ahead |
| Time of maximum | Time when the maximum occurred |

See chapter 3 for definitions and formulae.

## 2.2   Graphics box

The graphics part displays curves in random and continuous mode and reduced spectral measures in the other modes. The following x and y axes are relevant:

| | |
|---|---|
| Anticipation curve (red) | Time (top axis), time look-ahead $N$ (right axis). The numbers in scientific notation on the r.h.s. of the right axis show the distribution of the anticipation look-ahead over the range $[0, L]$ |
| Probability (black) | Time (top axis), probability (left axis.) |
| Variance (green) | Time (top axis), probability (left axis.). |
| Spectral measures | Position of eigenvalues in $[-\pi, \pi]$ (bottom axis), measure (left axis.) |

---

[4] Number and percent of all steps
[5] Number and percent of non-narrow
[6] Number and percent of positive
[7] Average taken over positive solutions
[8] Average taken over all steps



Quantum Anticipation Explorer

In continuous and random mode, in the upper part of the graphics box four bars (dashed lines) are drawn (in top-down order)

| | |
|---|---|
| Black bar | Positive solutions (25% in the graph on page 8) |
| Red bar | Solutions with zeros |
| Green bar | Singular spectra or solutions |
| Brown bar | Narrow reduced spectra |

The red and green bar are very sparse, usually, and in random mode likely to be empty.

## 2.3 Text box

This is multiple purpose box. Firstly, Quantum Anticipation Explorer displays certain error messages.

Secondly, in seek and single mode the spectral measure is displayed numerically:

| | |
|---|---|
| Index | The zero-based index of the eigenvalue in the original ordering. |
| Spectrum | The location of the eigenvalue in the interval $[-\pi, \pi]$. |
| Measure | The spectral measure of the eigenvalue. |

Thirdly, you can input custom spectra and spectral measures when selecting the "Prescribed" option in the Spectrum and/or Measure dropdown box. $d$ values separated by spaces, commas or tabs are required for each option chosen, i.e. $2d$ values to enter the whole spectral measure. The measure values must be non-negative and sum up to 1, with at most $1.0E-10$ error.

Caution: When moving the cursor into the box, the box will be cleared! Therefore, move the cursor out of the box before you start typing, and leave it there.

To re-use the values typed in, select the "Previous" option afterwards.

## 2.4 Numerical output part

The numerical output part on top displays values which are among those that can be saved (see 2.1). In seek an single mode the number of steps is always 1 and therefore the averages reflect the actual value.







## 3 Internals

This chapter explains what Quantum Anticipation Explorer does and how it works. We use here the same notation as in (Thomann, 2010) and refer to the results found there, but recall the definitions needed here. Therefore this chapter is self-contained.

### 3.1 Basic concepts

Quantum Anticipation Explorer works on a point spectrum of *dimension* $d$ given by mutually different real eigenvalues $\lambda_0 \cdots \lambda_{d-1}$ with non-negative spectral measure $\mu_0 \cdots \mu_{d-1}$ summing up to unity.

A spectral measure of dimension $d$ exhibits *orthogonal evolution* of order $L$ at time $T$ iff

(1) $\delta_k = \sum_{0 \leq n < d} \mu_n e^{-ik\lambda_n T}$ $(-L \leq k \leq L)$.

The state at time $T\Delta$ of a system of dimension $d$ exhibiting orthogonal evolution of order $L$ has inner product $\alpha_k = \sum_{0 \leq n < d} \mu_n e^{-i(k-\Delta)\lambda_n T}$ with the state at time $kT$. $\alpha_k$ $(-L \leq k \leq L)$ is the $k$-th *anticipation amplitude*, $p_k = |\alpha_k|^2$ the $k$-th *anticipation probability*. Then we define

| | |
|---|---|
| Anticipation look-ahead | $A = \sum_{k=0}^{L}(k-\Delta)p_k$ |
| Total anticipation probability | $P = \sum_{k=0}^{L} p_k$ |
| Variance | $V = \pi^{-1}\sqrt{\sum_{0 \leq n < d} \nu_n (\kappa_n - \overline{\kappa})^2}, \overline{\kappa} = \sum_{0 \leq n < d} \nu_n \kappa_n$ |
| Time look-ahead variable | $N = n$ iff $t = nT, 0 \leq n \leq L$ |

The anticipation look-ahead is related to the expectation of the time look-ahead variable measured at time $T\Delta$. The total anticipation probability is the probability to measure at time $T\Delta$ a value between 0 and $L$: $A = \text{Expectation}(\{N = n | measured\ at\ time\ T\Delta\} - P\Delta$.

$V$ is the variance of the reduced spectral measure. It is only included in this program to prepare for a possible future on the topics of (N. Margolus, 1998) does not matter in the context of quantum anticipation.

### 3.2 Main problem

The main problem solved by Quantum Anticipation Explorer is the following:

> Given order $L$, time $T$ and spectrum $\lambda_0 \cdots \lambda_{d-1}$, does there exist a non-negative measure $\mu_0 \cdots \mu_{d-1}$ such that the system exhibits orthogonal evolution of order $L$, i.e. satisfies equation system (1)?

Obviously, for general spectra $d \geq 2L + 1$, the *minimum dimension*, is required, otherwise the system is over-determined. See however (Thomann, 2010) for a thorough analysis. There are special spectra of lower dimension satisfying the equation system. The extreme case is the equidistant reduced spectrum of dimension $L + 1$ with equal measure.

If $d = 2L + 1$, then the system (1) has always a unique real-valued solution, due to its symmetry. Orthogonal evolution takes place iff this solution is non-negative.






If $d > 2L + 1$, then a manifold of dimension $2L + 1 - d$ of real solutions does exist. To determine if this manifold contains non-negative solutions leads to a Linear Programming Problem (LPP). We need the following definitions:

$$\Omega = \left(e^{-ik\lambda_n T}\right) (-L \leq k \leq L, 0 \leq n \leq 2L)$$
$$\Psi = \left(e^{-ik\lambda_n T}\right) (-L \leq k \leq L, 2L < n \leq d)$$
$$\omega = \text{Diag}\left(e^{-iL\lambda_n T}\right)(0 \leq n \leq 2L)$$
$$V = \Omega^T \omega$$
$$A = \Omega^{-1}\Psi$$
$$(1_i) = (\delta_{ik})(-L \leq k \leq L)$$
$$b = \Omega^{-1}(1_i)$$
$$\mu = (\mu_n)(0 \leq n \leq 2L)$$
$$I = \text{Diag}(1)(0 \leq n \leq 2L)$$

$\Omega$ is a square matrix, $\Psi$ a rectangular one, generally. $V$ (the product of the transpose of $\Omega$ with $\omega$, a diagonal matrix) is the Vandermonde matrix (Aitken, 1956). $(1_i)$ is the i-th unit vector, $I$ the identity matrix.

With this notation (1) is equivalent to $(\Omega|\Psi)\mu = (1_i)$. Provided that $\left(e^{-i\lambda_n T}\right) (0 \leq n \leq 2L)$ are mutually different, the relationship to the Vandermondian implies that $\Omega$ is non-singular. Its inverse is calculated in $O(L^2)$ operations using the Parker algorithm (Parker, 1964), which is further improved utilizing the row symmetry of $\Omega$ and the special form of its matrix elements. Multiplying (1) by this inverse yields the equivalent system

(2) $(I|A)\mu = b$

The following lemma implies that $A$ and $b$ are purely real, because $\Omega$ and $\Psi$ are row-symmetric.

<u>Lemma 1: Row- and column-symmetric matrices and symmetric vectors</u>
Call a $n-$vector $v$ symmetric, iff $v_i = \overline{v_{n-i}}$, a matrix row-symmetric (column-symmetric) iff all column (row) vectors are symmetric. Then

a) The product of a column-symmetric matrix with a symmetric vector is real.
b) The inverse of a row- (column-) symmetric matrix is column- (row-) symmetric.

Proof: The first part holds because the sum contains to each product also its complex conjugate. To obtain the second part, notice that the space of symmetric $n-$vectors, $S$, is isomorphic with $\mathbb{R}^n$. Any non-singular column-symmetric matrix is an isomorphism $S \to \mathbb{R}^n$. Thus its inverse is an isomorphism from $\mathbb{R}^n \to S$. Particularly, it maps $(1_i)$ to $S$, for any $i$. Thus its columns are symmetric, implying the lemma.
∎

By (Ferguson, 2008) (2) is an LPP in canonical form with $2L + 1$ constraints. The first $2L + 1$ elements of $\mu$ are slack variables, the remaining $d - 2L - 1$ elements structural variables.

Feasible solutions are those non-negative $\mu$ satisfying (2). As for, $k = 0$, (1) requires $\sum \mu_n = 1$ which implies $\sum_{n \geq 2L} \mu_n \leq 1$, the problem is either infeasible or bounded feasible.

In all modes, if the measure type is "Optimum", then Quantum Anticipation Explorer determines whether or not a feasible solution exists using the Simplex Algorithm. If one exists, then it proceeds to find one maximizing the anticipation look-ahead. As this quantity is a qu-adratic function of the spectral measure, linear programming alone does not yield it. We apply the following algorithm:

0. Find an initial feasible solution.
1. Starting from the current solution, $\mu$, determine the gradient $\nabla$ of the anticipation look-ahead.
2. Taking the gradient[9] as cost function, determine the optimum solution, $\mu'$, using the Simplex algorithm.
3. If the new optimum solution improves the anticipation look-ahead by more then $0.001$ then go to 1, else output the optimum solution.

The simplex algorithm, using a linear cost function, yields solutions lying in the corners of a simplex on a hyperplane of dimension $2L + 1$ in $d$-space. The above algorithm finds the optimum based on the premise that the anticipation look-ahead is a non-negative quadratic function. The premise is satisfied because we restrict the location of measurement to negative values.

The positive solutions of the main problem coincide with the feasible solution of the above linear programming problem, i.e. they are lying in the said simplex. If a point is not optimal, then better points are lying in the direction of the projection of the gradient onto the simplex.

Consider now two successive corners visited by our algorithm. We claim that on the edge between these two corners the optimum must lie at the end point, i.e. the last visited corner.

For proof, first notice that, by the positivity of the anticipation look-ahead, its equipotential surfaces are ellipsoids centered at the origin, and the gradient points outward.

If the claim were wrong, then there were an intermediate point on the edge between the last and second-last corner visited where the gradient were orthogonal to this edge. The edge would thus be a tangent to the ellipsoid, touching it in the said intermediate point, while the endpoints were lying on ellipsoids with larger axes. As the gradients point outward, the starting and the end point thus had larger anticipation look-ahead than the intermediate point. But then the gradient at the second-last corner would point away from this point and the last corner. Therefore the latter had never been visited by the simplex algorithm. q.e.d.

The corners of the simplex are characterized by the condition that all structural variables are zero. Therefore, if a feasible solution exists, then its *non-zero dimension* (i.e. the number of eigenvalues with non-zero[10] measure) is less or equal the minimum dimension, $2L + 1$.

This proves

### Theorem 1: Optimal solutions
The positive solutions of the main problem are lying in a simplex, optimal solutions in its corners. The non-zero dimension of optimal solutions is less or equal the minimum dimension, $2L + 1$. They are found by the above algorithm.

∎

---

[9] By equation (2), the $2L + 1$ slack variables are a function of the $d - 2L - 1$ structural variables. The gradient is evaluated in terms of the latter ones, and thus projected on the simplex.
[10] The program tests for $> 10^{-4}$.

However, solutions with higher non-zero dimension are easily found by convex interpolation between successive solutions. It is obvious that convex sums of solutions of (1) are again solutions.

For numerical reasons, the algorithm may not always yield accurate results. If either the l.h.s and r.h.s. of (1) are different by more than 0.001 in Eucledian norm, or the total anticipation probability $P$ is negative of greater 1.001, then the solution is called *singular*.

Non-singular solutions of the main problem are called *positive*.

## 3.3   Spectra

This section introduces some special properties and treatments of spectra.

Due to the periodicity of the exponential function, $\lambda_n T$ can be replaced in (1) by the *reduced spectrum* $\kappa_n = \lambda_n T \mod 2\pi$. Different eigenvalues in the total spectrum may degenerate to the same value in the reduced spectrum.

The *reduced spectral measure* $\nu_n$ is easily obtained from $\mu$, cumulating the measure if two or more eigenvalues degenerate to the same value in the reduced spectrum.

It is the reduced spectrum that is displayed in seek and single mode.

At any time $T$, the first thing done is the determination of the reduced spectrum. The reduced spectrum is then checked and cleaned up to discover possible singularities and to take the necessary provisions:

Firstly, whenever in the reduced spectrum two eigenvalues have a distance of less than $10^{-6}$, then the spectrum is considered *degenerate* and one of them is removed. The number of the remaining eigenvalues is the *actual dimension*. If it is less than $2L + 1$, then the system (1) cannot be solved, the spectrum is *singular*.

Secondly, if the problem is non-singular, then reduced eigenvalues which are close to others are moved to the end of the list. This assures that $\Omega$ is not only non-singular but also as well-conditioned as possible.

Next the *spectral width*, a simple property necessary for the existence of positive solutions is checked, as stated in lemma 2 below.

Define the spectral width at time $t$ as the maximum distance between any two eigenvalues adjacent on the unit circle in the reduced spectrum at that time. I.e. take the reduced spectrum, then map the eigenvalues on the unit circle, and consider the distances between adjacent values.

Lemma 2:

a) Necessary and sufficient for the existence of a positive solution of equation (1) for $L = 1$ is that the spectral width at time $T$ be less or equal $\pi$.
b) Necessary for the existence of a positive solution of equation (1) for $L > 1$ is that the spectral width at time $nT$ be less or equal $\frac{1}{L+1}$.

Proof: The first part follows from the observation that, if all eigenvalues are lying in the right half-plane, then all cosines are non-negative, therefore cannot sum up to zero. The proof of the second part is more involved and will be provided in another publication.

Spectra meeting this criterion are deemed *non-narrow* and are further processed. The spectral measure is determined according to the option selected in the Measure dropdown menu. If this measure solves the main problem, then the solution is deemed *positive*.

Given a positive solution, the number of eigenvalues whose spectral measure exceeds 0.0001 is its *non-zero dimension*. The number of solutions with non-zero dimension less than $2L + 1$ and the average value of the non-zero dimension part of the numerical output.

## 4      Appetizers

- Assess the anticipation strength:
    - Select H-Atom spectrum, Optimum measure and Continuous mode, click the From and the To label and set the order to 1 and the dimension to 3. Now press Go. Notice the average and maximum anticipation. Press Show Max to see the spectral measure at maximum anticipation time. It is close to the equal distribution. Notice also the anticipation probability values.
    - Increase the dimension and repeat the evaluation. The average anticipation converges to 1, the maximum to 1.5.
    - Increase the order and the time interval, and repeat the evaluation at increasing dimensions. Notice the numbers at the r.h.s. of the diagram, displaying e.g. at order 2 and dimension 11 64% anticipation between 1 and 2, 36% between 0 and 1.
- Find the smallest T where Anticipation occurs in the H-Atom:
    - Select H-Atom spectrum, Optimum measure and Seek Positive mode, and click the From label and the To label. Set the step size to 0.0001, the order to 1 and the dimension to 3 and press the ">" button. You get the equal distribution at time t = 9/16. You can check this by entering numerator 9 and denominator 16, double-clicking into the From box, selecting Single mode and pressing the Go button. Notice that T exactly answers the question of (N. Margolus, 1998).
    - Double-click the From button again, increase the dimension to 4 and press ">". As you increase the dimension and reduce the step size down to 0.00001, the time of first positive measure converges to 0.5.
    - Now set the dimension to 30, increase the order by one, press ">", again increase the order and press ">", a.s.f..
- Find solutions of dimension less than $2L + 1$:
    - Select the H-Atom spectrum, optimum measure and "Seek dim. chg." mode. Press ">". At order 1, dimension 3 and step size 0.0001 the evaluation stops at time 0.5625, like the previous one. Notice that the non-zero dimension is 2. Now press ">" again: At time 0.5626 the non-zero dimension is 3. It changes again to 2 at time 1.6875. Notice that the minimum dimension is $L + 1$, realized by equidistant spectrum.
    - Repeat this with other orders and higher dimensions.
    - Repeat this with random spectra.
- Verify the periodicity of the A-Atom spectrum:
    - Select H-Atom spectrum, Optimum measure and Continuous mode, and set the order to 1 and the dimension to 3. Click the From label and enter 144 in the To box. Press Go. The left and the right half of the diagram are identical. The period is $72\,Pi = 2 \cdot 2 \cdot 3 \cdot 3 \cdot 2Pi$.
- See how frequently anticipation occurs in atomic spectra:
    - Select H-Atom spectrum, Optimum measure, Continuous mode, click the From and the To label, then press Go. Now increase again the dimension. At order 1 and dimension 30, of the 72000 steps 70089 yield a non-narrow spectrum, 70088 are positive, almost 100%! You see, the criterion of lemma 2 is fairly efficient, and anticipation fairly likely. Notice that 206 spectra are degenerate, 1 is singular. Also, 61 zeros occur.
    - Still at order 1 and dimension 30 change the mode to Random and press Go. Notice the change of degenerate, singular an zeros. This effect is even more

    

       drastic at minimum dimension (3). Many singularities occur in continuous mode, due to the special numerical properties of the eigenvalues -1, -1/4 and -1/9 of the H-Atom spectrum. In random mode these are less likely to occur.
- Verify that anticipation is ubiquitous:
    - Select random spectrum and optimum measure of various orders and increasing dimensions, and evaluate anticipation in some time interval in continuous mode (random mode does not make much sense in randomly chosen spectra). Notice that, at sufficiently large dimensions, the average anticipation is always greater than half the order, and the maximum anticipation close to the order.
    - Every time you press the Go button, a new random spectrum will be chosen. If you wish to continue the evaluation of some spectrum, select the "Previous" option in the spectrum dropdown menu.
- Analyze your favorite spectra and spectral measures using the Prescribed option. Notice that dense and continuous spectra can be approximated by high-dimensional point spectra (see theorem 3 of (Thomann, 2010)).
- Export numerical results and graphics using the save button.

 17/18